\documentclass[%
 reprint,
 superscriptaddress,
 amsmath,amssymb,
 aps,
 prl,
 floatfix,
]{revtex4-2}

\usepackage{graphicx}
\graphicspath{{.}}
\usepackage{dcolumn}
\usepackage{bm}
\usepackage{color} 
\usepackage[hypertexnames=false]{hyperref}
\definecolor{OURblue}{RGB}{46, 48, 146}
\hypersetup{%
	colorlinks=true,
	linkcolor=OURblue,
	citecolor=OURblue,
	urlcolor=OURblue 
}
\usepackage{siunitx}
\usepackage[T1]{fontenc}
\usepackage{ae}

\newcommand{\Dline}[1]{\ensuremath{D_{#1}}}

\newcommand{\eqselfbroadening}{\ref{eq:self-broadening}}
\newcommand{\eqdipoledipoleshift}{\ref{eq:dipole-dipole_shift}}
\newcommand{\figlongCell}{\ref{fig:longCell}}
\newcommand{\figshortCell}{\ref{fig:shortCell}}
\newcommand{\figDoneDtwo}{\ref{fig:broadening_shift}}

\begin{document}

\preprint{2021_LIAD_arXiv}

\title{Transient Density-Induced Dipolar Interactions in a Thin Vapor Cell}
\author{Florian Christaller}
\author{Max M\"ausezahl}
\author{Felix Moumtsilis}
\affiliation{%
	5.~Physikalisches Institut and Center for Integrated Quantum Science and Technology,\\
	Universit\"at Stuttgart, Pfaffenwaldring 57, 70569 Stuttgart, Germany
}%
\author{Annika Belz}
\affiliation{%
	5.~Physikalisches Institut and Center for Integrated Quantum Science and Technology,\\
	Universit\"at Stuttgart, Pfaffenwaldring 57, 70569 Stuttgart, Germany
}%
\affiliation{%
	Department of Physics, Joint Quantum Centre (JQC) Durham-Newcastle, Durham University, 
	South Road, Durham, DH1 3LE, United Kingdom
}%
\author{Harald K\"ubler}
\affiliation{%
	5.~Physikalisches Institut and Center for Integrated Quantum Science and Technology,\\
	Universit\"at Stuttgart, Pfaffenwaldring 57, 70569 Stuttgart, Germany
}%
\author{Hadiseh Alaeian}
\affiliation{%
	Elmore Family School of Electrical and Computer Engineering, Department of Physics and Astronomy, 
	Purdue Quantum Science and Engineering Institute, Purdue University, West Lafayette, 
	Indiana 47907, USA
}%
\author{Charles S. Adams}
\affiliation{%
	Department of Physics, Joint Quantum Centre (JQC) Durham-Newcastle, Durham University, 
	South Road, Durham, DH1 3LE, United Kingdom
}%
\author{Robert L\"ow}
\author{Tilman Pfau}%
\email{t.pfau@physik.uni-stuttgart.de}
\affiliation{%
	5.~Physikalisches Institut and Center for Integrated Quantum Science and Technology,\\
	Universit\"at Stuttgart, Pfaffenwaldring 57, 70569 Stuttgart, Germany
}%
\date{\today}
\begin{abstract}
	We exploit the effect of light-induced atomic desorption to produce high atomic densities 
	($n\gg k^3$) in a rubidium vapor cell. An intense 
	off-resonant laser is pulsed for roughly one nanosecond on a micrometer-sized sapphire-coated cell, 
	which results in the desorption of atomic clouds from both internal surfaces. We probe 
	the transient atomic density evolution by time-resolved absorption spectroscopy.
	With a temporal resolution of $\approx\SI{1}{\nano\second}$, we measure the broadening and line shift 
	of the atomic resonances. Both broadening and line shift are attributed to dipole-dipole interactions.
	This fast switching of the atomic density and dipolar interactions could be the basis for future quantum 
	devices based on the excitation blockade.
\end{abstract}
%
\maketitle
%
%
The effect of dipole-dipole interactions in optical media becomes important when the density $n$ is 
significantly larger than the wave number cubed $k^3$ of the interaction light field. 
Entering this regime leads to interesting nonlinear effects such as an excitation 
blockade \cite{Lukin2001}, nonclassical photon scattering \cite{Williamson2020}, 
self-broadening (collisional broadening) \cite{Lewis1980}, and the collective 
Lamb shift \cite{Lamb1947,Friedberg1973}. 
\\
Dipole-dipole interactions are observable 
in steady-state experiments performed in thin alkali vapor cells \cite{Weller2011,Keaveney2012,Peyrot2018},
where the cells are heated to temperatures above $\SI{300}{\degreeCelsius}$.
Dipolar broadening effects were previously observed to be independent 
of the system geometry, while the line shift depends on the dimensionality of the system, 
as investigated in a 2D model \cite{Peyrot2018,Dobbertin2020}. It is however not straightforward 
to prepare high densities with alkali vapors \cite{Lorenz2009}.
\\
One technique to increase the atomic density is light-induced atomic desorption 
(LIAD) \cite{Gozzini1993,Meucci1994,Alexandrov2002,Rebilas2009,Petrov2017,Talker2021} or
light-activated dispensers \cite{Griffin2005}. LIAD is commonly used for loading 
magneto-optical traps \cite{Anderson2001,Atutov2003,Klempt2006}
and has been studied in confined geometries
like photonic-band gap fibers \cite{Ghosh2006,Slepkov2008} or porous samples \cite{Burchianti2006}. 
However, the application of pulsed LIAD for fast switching of dipole-dipole interactions 
is so far unexplored.
\\
In our pulsed LIAD setup, we can switch the atomic density in the nanosecond domain, which allows 
one to study the dipole-dipole interaction on a timescale faster 
than the natural atomic lifetime. 
This fast density switching has been already used in our group to realize an on-demand room-temperature 
single-photon source based on the Rydberg blockade \cite{Ripka2018}.
In this work, we study the dipolar interaction for the two transitions \Dline{1}: 
$5S_{1/2} \to 5P_{1/2}$ and \Dline{2}: $5S_{1/2} \to 5P_{3/2}$ of rubidium 
with different transition dipole moments.
\\
We first describe our LIAD measurement results in a thicker part 
of the cell [cell thickness $L = \SI{6.24 \pm 0.07}{\micro\metre}$] at a low density ($nk^{-3}\approx1$). 
This measurement is used as the basis to set up a model for the velocity and density distribution 
of the desorbed atoms. Then, we focus on a thinner part of the cell [$L = \SI{0.78 \pm 0.02}{\micro\metre}$], 
where we can study transient density-dependent dipolar interactions at a 
high density (up to $nk^{-3}\approx100$). 
To this end, we compare two transitions (\Dline{1} and \Dline{2} transition of rubidium) with different 
transition dipole moments to investigate their influences on the dipole-dipole interaction 
in a quasi-2D geometry ($L \approx \lambda_\text{probe}$).
\par
%
%
\begin{figure}[t]
	\includegraphics[width=\linewidth]{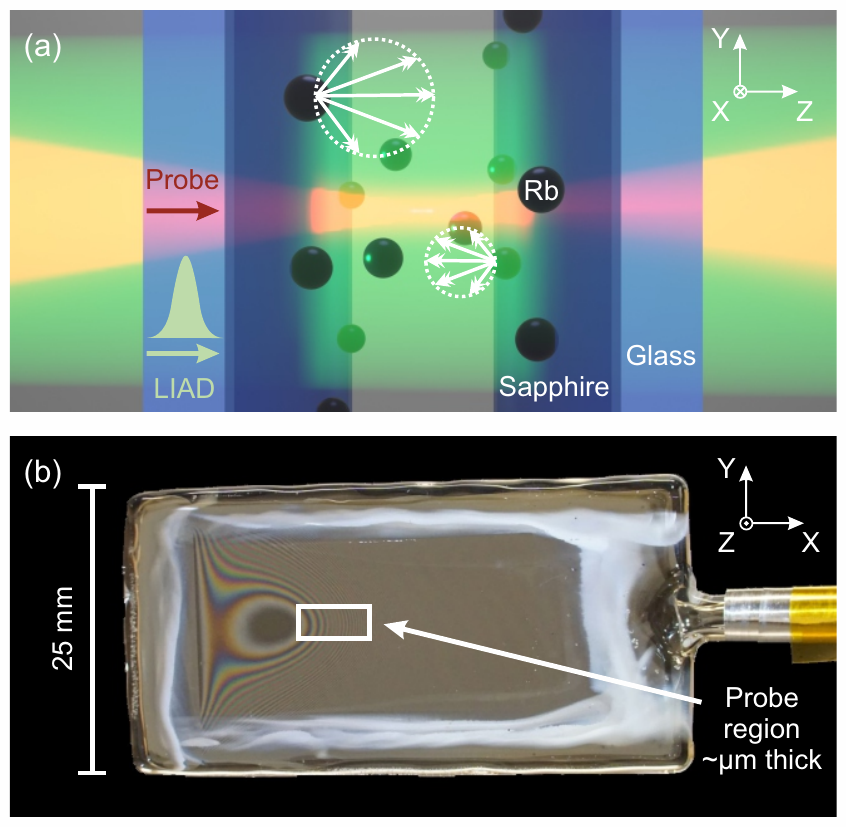}
	\caption{\label{fig:principle} 
		(a) Illustration of the LIAD process. A green laser pulse (LIAD pulse) 
		desorbs rubidium (Rb) atoms, which are adsorbed on the inner, sapphire-coated surface of the glass cell. 
		The desorbed atoms are emitted with a certain velocity distribution into the cell volume 
		from both sides of the cell, where they are probed with a red laser (probe laser).
		(b) Photo of the wedge-shaped micrometer-sized cell with interference fringes (Newton's rings). 
		The cell thickness in the probe region is determined interferometrically and ranges 
		from $\num{0.78 \pm 0.02}$ to $\SI{6.24 \pm 0.07}{\micro\meter}$ from left to right.
	}
\end{figure}
\begin{figure}[b]
	\includegraphics[width=\linewidth]{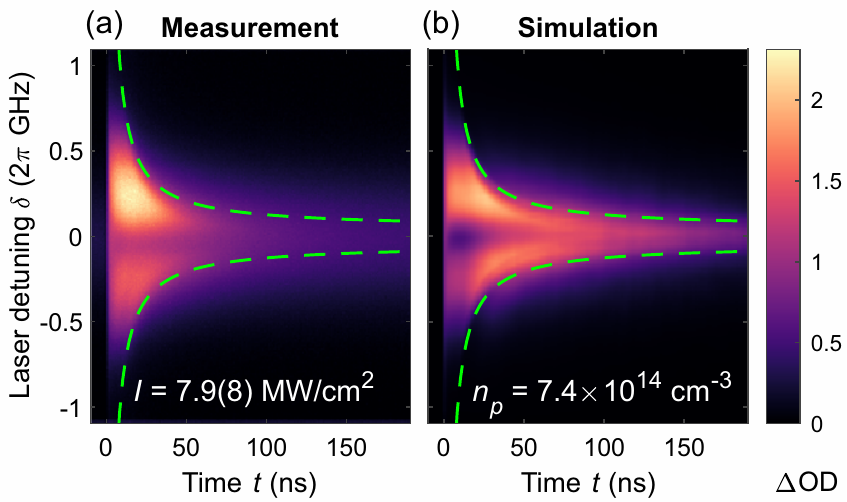}
	\caption{\label{fig:longCell} 
		(a) Measured $\Delta \text{OD}$ map of the $^{85}$Rb $F_g=2$ transition 
		of the \Dline{2} line. After the atoms, depending on their detuning ($z$ velocity), hit 
		the other cell wall the signal decreases. The $\text{TOF}(\delta)$ is shown with two dashed green lines. 
		The intensity $I$ is the peak intensity of the LIAD pulse. The cell thickness is 
		$L = \SI{6.24 \pm 0.07}{\micro\metre}$ and the reservoir temperature is 
		$T_\text{res} \approx \SI{140}{\degreeCelsius}$.
		(b) Simulated $\Delta \text{OD}$ map of desorbed atoms. The simulation parameters 
		are adapted to the measurement. The $\text{TOF}(\delta)$ curves (dashed green lines) are plotted, too. 
		The indicated $n_p$ is the peak density at $\SI{2}{\nano\second}$.
	}
\end{figure}

We use a self-made, wedge-shaped, micrometer-sized cell with an attached reservoir tube 
filled with rubidium ($72\%$ $^{85}$Rb, $28\%$ $^{87}$Rb), shown in Fig.~\ref{fig:principle}(b). 
By heating the cell independently from the reservoir, we can produce a certain rubidium coverage 
on the cell walls and a vapor pressure in the cell with a comparably small background density 
$n \approx \SI[parse-numbers = false]{10^{14}}{\per\centi\metre\cubed}$ 
(reservoir temperature $T_\text{res}\approx\SI{180}{\degreeCelsius}$). 
In our setup, the pulsed LIAD laser and the probe laser are aligned collinearly in front 
of the cell [Fig.~\ref{fig:principle}(a)]. The pulsed LIAD laser at $\SI{532}{\nano\meter}$ 
has a pulse length of $\SI{1.1\pm0.1}{\nano\second}$ (FWHM) and a repetition rate 
of $\SI{50}{\kilo\hertz}$. This off-resonant pulse leads to the desorption of rubidium atoms 
bound to the sapphire-coated glass surface. The amount of desorbed atoms depends 
on the peak intensity $I$ of the LIAD pulse. 
We probe the desorbed atoms at $\SI{795}{\nano\meter}$ (\Dline{1} transition of Rb) or 
$\SI{780}{\nano\meter}$ (\Dline{2} transition of Rb). Both probe lasers have an intensity well below 
the resonant \Dline{2} saturation intensity $I_\text{probe} < 0.01 I_\text{sat,\Dline{2}}$. 
The measured Gaussian beam waist radius ($1/e^2$) 
of the probe laser is $w_\text{probe} = \SI{2.0 \pm 0.2}{\micro\meter}$, while 
the LIAD laser has a waist radius of $w_\text{LIAD} = \SI{13.7 \pm 0.1}{\micro\meter}$. 
We measure the transmitted photons with a single-photon counting module, 
which is read out by a time tagger module. 
Our wedge-shaped cell has a point of contact of the cell walls, which can be seen in 
Fig.~\ref{fig:principle}(b) as a dark circle. To the right of this point lies the probe region where
the cell is less than $\SI{10}{\micro\metre}$ thick. The local cell thickness 
can be directly determined interferometrically by counting Newton's rings.
\\
During the measurement, the probe laser is scanned over the \Dline{1} or \Dline{2} transition 
at a slow frequency of $\SI{11}{\hertz}$. At the same time, the LIAD laser sends pulses with 
a high repetition rate ($\SI{50}{\kilo\hertz}$) into the cell. 
We take full scans of the probe detuning $\delta$ at different times $t$ 
after the LIAD pulse (see Supplemental Material \cite{SupplMaterial}\nocite{Sekiguchi2017,Loudon2000,SteckRb,Knudsen1934,Sibalic2017}). 
The time-resolved transmission $T(t,\delta)$ of the probe laser is used to calculate 
the change of the optical depth $\Delta \text{OD}$. For every detuning, the transmission 
before the LIAD pulse is used as the background signal $T_0(t<0,\delta)$, which is used 
to calculate $\Delta \text{OD}(t,\delta) = \ln{(T_0/T)}$. Thereby the optical depth caused 
by the background vapor pressure is subtracted. A map of the time- and 
detuning-resolved $\Delta \text{OD}$ is shown in Fig.~\ref{fig:longCell}(a). 
At $t=\SI{0}{\nano\second}$ the LIAD pulse hits the cell and increases the optical depth. 
The time resolution of the measurements is limited by the time jitter of the LIAD pulse 
($\SI{500}{\pico\second}$) and the single-photon counting module ($\SI{350}{\pico\second}$).
\par
%
%
First, we focus on a thicker part of the cell [$L = \SI{6.24 \pm 0.07}{\micro\metre}$], where we measure 
a time- and detuning-resolved $\Delta \text{OD}$ map of the $^{85}$Rb $F_g=2$ transition 
of the \Dline{2} line [Fig.~\ref{fig:longCell}(a)]. The transition is defined by the total angular 
quantum number $F_g$ of the ground state $g$, while the total hyperfine splitting $\delta_\text{hfs}$ 
of the excited state cannot be resolved due to transient and Doppler broadening.
The atoms moving in the laser propagation direction, 
originating from the entry facet of the cell, are probed at blue detunings $\delta > 0$. 
A second group of atoms, originating from the exit facet, is visible and probed at red detunings 
$\delta < 0$. The signal is higher for the atoms moving in the laser propagation direction. 
This asymmetry is not anticipated, but might originate from differences in the surface properties
as it was observed in other experiments, i.e., depending on the coating \cite{Ghosh2006}. 
We checked this hypothesis by rotating the cell by \ang{180}, which led to a roughly inverted asymmetry.
The darker region around zero detuning shows that fewer atoms 
with low $z$ velocity are desorbed. In total, we measure two atom clouds moving toward 
the opposite cell walls. The high $\Delta \text{OD}$ value in the first nanoseconds is caused 
by a high atomic density and decreases over time. The $\Delta \text{OD}$
signal equilibrates to zero before the next LIAD pulse arrives. The dashed green lines indicate 
the time-of-flight curves, after which the atoms with a certain detuning hit the other cell wall 
according to $\text{TOF}(\delta) = L k / |\delta|$, 
with $\delta = \boldsymbol{k}\cdot \boldsymbol{v} \pm \delta_\text{hfs}/2 = k v_z \pm \delta_\text{hfs}/2$, respecting the hyperfine splitting of the excited state. There, $\boldsymbol{k}$ is the wave vector 
of the probe beam, which is parallel to the $z$ axis, $k=|\boldsymbol{k}|$ is the wave number 
of the probe beam, $\boldsymbol{v}$ is the velocity of the atom, and $v_z$ is the $z$ component 
of the velocity. We observe distinct signal wings beyond the respective time-of-flight curves
indicating potential re-emission of atoms after arriving at the opposite cell wall.
\\
Using this measurement as a reference, we develop a kinematic model and run a Monte Carlo simulation 
of atoms flying through a cell and interacting with 
the probe laser (see Supplemental Material \cite{SupplMaterial}). The idea is to model 
the velocity distribution of desorbed atoms and to estimate the local density during the simulation.
\\ 
In the model, the local and temporal desorption-rate scales linearly with the intensity of the LIAD pulse. 
For the velocity distribution we assume 
$f(v,\varphi,\theta) = \left[4/(\sqrt{\pi}a^3)v^2 \exp{(-v^2/a^2)}\right] \cos{(\theta)}$ with the parameter $a$ 
and the speed $v=|\boldsymbol{v}|$. The azimuthal angle $\varphi$ is uniformly distributed, while the 
polar angle $\theta$ is distributed according to the $\cos{(\theta)}$-Knudsen law \cite{Comsa1985}.
This simple distribution leads to a good qualitative agreement between measurement 
[Fig.~\ref{fig:longCell}(a)] and simulation [Fig.~\ref{fig:longCell}(b)]. We also assume 
that the atoms are desorbed only during the LIAD pulse and that there is no thermal desorption 
after the pulse, which is in good agreement with our measurement. 
Since no other mechanism (i.e., through natural- or transit-broadening, which are also included in the model) 
reproduces the signal wings beyond the time-of-flight curves (dashed green lines in Fig.~\ref{fig:longCell}),
they might occur because of re-emissions from the surfaces after bombardment with the initial atom clouds.
To get better agreement, an instant re-emission probability of $\SI{84}{\percent}$ is included in the
kinematic model.
\\
The remaining discrepancies between measurement and simulation 
can originate from an inadequate velocity distribution model, 
intricate re-emission properties, additional decay mechanisms, the neglected Gaussian intensity distribution 
of the probe beam, and the use of the steady-state cross section of the atoms at all the times. 
Nevertheless, with the overall acceptable agreement between measurement and simulation 
we obtain a time- and $z$-dependent simulated local density, 
which shows that the desorbed atoms are initially in two flat, ``pancakelike'' clouds 
with an initial thickness well below the wavelength of the probe laser, rendering this into 
a 2D geometry (see Supplemental Material \cite{SupplMaterial}).
\par
%
%
\begin{figure}[t]
	\includegraphics[width=\linewidth]{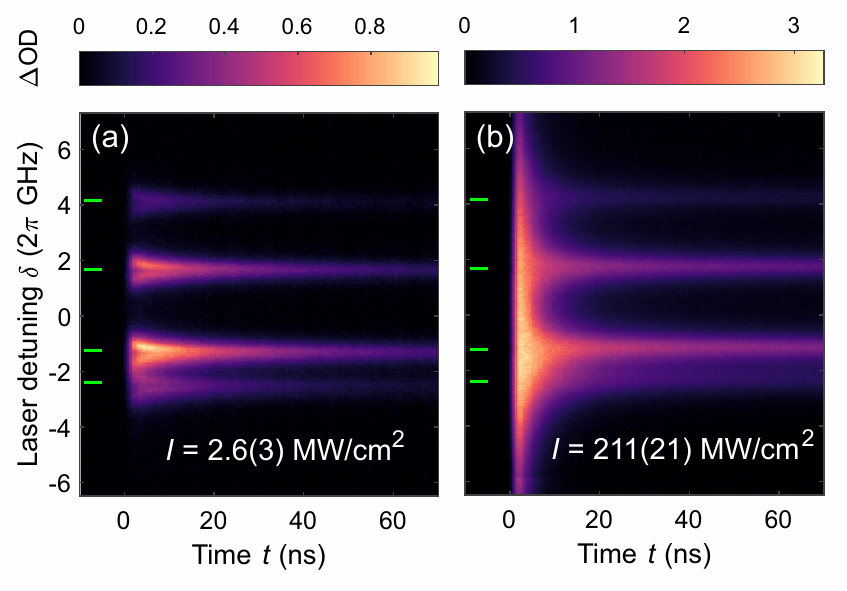}
	\caption{\label{fig:shortCell} 
		Measurements of the time- and detuning-resolved $\Delta \text{OD}$ 
		for low density (a) and high density (b) at the \Dline{2} transition.
		There is a broadening and line shift of the spectrum in the first few nanoseconds for the high-density case. 
		The measured broadening is mainly attributed to the density dependent self-broadening 
		($\Gamma_\text{self}$). The line shift occurs due to the dipole-dipole shift ($\Delta_\text{dd}$,
		more visible in the slices presented in the Supplemental Material \cite{SupplMaterial}).
		The four green markers indicate the ground state hyperfine splitting 
		of the two isotopes of rubidium, respectively. 
		The intensities $I$ are the peak intensities of the LIAD pulse, the cell thickness is 
		$L = \SI{0.78 \pm 0.02}{\micro\metre}$, and the reservoir temperature is 
		$T_\text{res} \approx \SI{180}{\degreeCelsius}$.
	}
\end{figure}
To investigate high-density regimes, we use a thinner part of the cell, 
as the background optical depth and the detection limit of the single-photon counting module 
are limiting the measurement in the thicker part of the cell. We perform measurements 
at a cell thickness of $L = \SI{0.78 \pm 0.02}{\micro\metre}$ at low 
and high atomic densities using the \Dline{2} transition as shown 
in Figs.~\ref{fig:shortCell}(a) and (b), respectively. Our measurements 
are in a regime where the total number of desorbed atoms per pulse 
monotonically increases with the peak intensity of the LIAD pulse 
(see Supplemental Material \cite{SupplMaterial}). The low-density measurement 
corresponds to a peak intensity of 
$I = \SI{2.6\pm0.3}{\mega\watt\per\square\centi\metre}$, 
while the high-density case corresponds to 
$I = \SI{211\pm21}{\mega\watt\per\square\centi\metre}$. 
\\
There is a broadening and line shift of the \Dline{2} hyperfine transitions present 
in Fig.~\ref{fig:shortCell}(b), which we attribute to density-dependent dipole-dipole interactions. 
The four peaks correspond to the ground state hyperfine splitting of the two isotopes of rubidium, 
contributing to the signal, while the hyperfine splitting of the excited state can not be resolved. 
The density-dependent self-broadening \cite{Lewis1980,Weller2011} in the steady-state regime 
was predicted to be 
\begin{equation}
	\Gamma_\text{self} = \beta_{i} n = \frac{2}{3\hbar\epsilon_0} \sqrt{\frac{g_g}{g_e}} 
	d_J^2 n~,
	\label{eq:self-broadening}
\end{equation}
where $\beta_{i}$ is the self-broadening coefficient, $i$ enumerates the \Dline{1} or \Dline{2} transition, 
$\hbar$ is the reduced Planck constant, $\epsilon_0$ is the vacuum permittivity, $g_g$ and $g_e$ 
are the multiplicities (depending on the quantum number $J$) of the ground and excited state respectively, $d_J$ 
is the total reduced dipole matrix element, and $n$ is the atomic density. In the high-density regime 
in Fig.~\ref{fig:shortCell}(b) 
we observe a self-broadening of $\Gamma_\text{self} \approx 590 \Gamma_0$ at $t=\SI{2}{\nano\second}$, 
where $\Gamma_0 \approx 2\pi \times \SI{6.07}{\mega\hertz}$~\cite{Volz1996} is the natural decay rate 
of the \Dline{2} transition.
\\
Similarly, we compare the line shift, observed in our measurements, to 
the steady-state dipole-dipole shift \cite{Friedberg1973,Keaveney2012}, 
which was predicted to be
\begin{equation}
	\Delta_\text{dd} = -\left|\Delta_\text{LL}\right| + \frac{3}{4} \left|\Delta_\text{LL}\right| 
	\left(1-\frac{\sin{2kL}}{2kL} \right)~,
	\label{eq:dipole-dipole_shift}
\end{equation}
with $\Delta_\text{LL}$ being the Lorentz-Lorenz shift and $L$ being the cloud thickness. 
This thickness dependency is a cavity-induced correction, also known as the collective Lamb shift. 
The Lorentz-Lorenz shift \cite{Friedberg1973,Keaveney2012}, in turn, is density-dependent 
and can be written as
\begin{equation}
	\Delta_\text{LL} = -\frac{1}{3\hbar\epsilon_0} d_J^2 n~.
	\label{eq:Lorentz-Lorenz_shift}
\end{equation}
As our cell thickness is $L \approx \lambda$, the second term of the dipole-dipole shift 
has a significant effect on the line shift and reduces the dipole-dipole effect to 
$\Delta_\text{dd} \approx -\frac{1}{4} \left|\Delta_\text{LL}\right|$. In the 
high-density measurement in Fig.~\ref{fig:shortCell}(b) this corresponds to 
a value of $\Delta_\text{dd} \approx -80 \Gamma_0$ 
(redshift) at $t=\SI{2}{\nano\second}$. Additionally, we can observe that the transient 
density-dependent effects occur on a timescale of a few nanoseconds, which is faster 
than the natural lifetime of the \Dline{2} transition ($\SI{26.2}{\nano\second}$)~\cite{Volz1996}.
\par
%
%
\begin{figure}[b]
	\includegraphics[width=\linewidth]{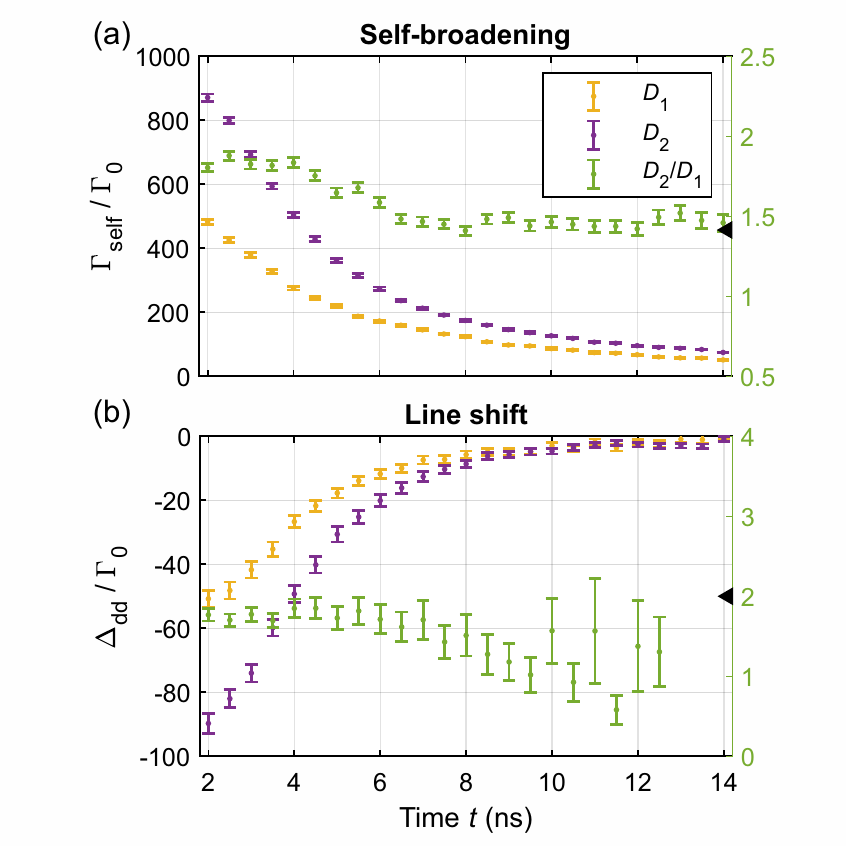}
	\caption{\label{fig:broadening_shift}
		(a) Time-dependent self-broadening $\Gamma_\text{self}$ of the Rb spectra for the 
		\Dline{1} (yellow data) and \Dline{2} (purple data) transition. The experiments 
		are performed under identical conditions, except the probe laser wavelength. 
		The ratio (green data) of the self-broadening of the two transitions 
		approaches the theoretical steady-state ratio for large $t$ (black triangle).
		Vertical error bars for $t<\SI{8}{\nano\second}$ are likely underestimated due to a systematic effect,
		and horizontal precision is bandwidth limited by jitter.
		(b) Time-dependent line shift $\Delta_\text{dd}$ of the \Dline{1} and \Dline{2} transition.
		The ratio of the line shift, which is close to the theoretical value, 
		has an increasing error for increasing time, and 
		therefore the last three data points were omitted.
		The peak intensity of the LIAD pulse is $I = \SI{317\pm32}{\mega\watt\per\square\centi\metre}$. 
		The cell thickness is $L = \SI{0.78 \pm 0.02}{\micro\metre}$ and the reservoir temperature is 
		$T_\text{res} \approx \SI{180}{\degreeCelsius}$.
	}
\end{figure}
To further investigate the dipole-dipole origin of the observed interaction, we compare 
the transient evolution of the self-broadening and line shift at the \Dline{1} and \Dline{2} 
transition of rubidium. We fit both measured data with a steady-state electric susceptibility model 
at each time step, using the software \textsc{ElecSus} \cite{Zentile2015}. The fits to 
the individual time-resolved spectra show a $<\SI{6}{\percent}$ 
overall normalized root-mean-square deviation and result in 
the self-broadening and line shift shown in 
Figs.~\ref{fig:broadening_shift}(a) and (b), respectively. Note that in the first $\SI{2}{\nano\second}$ 
we cannot properly fit the data to this model, so we exclude these data points. 
The error bars represent the $1\sigma$ standard fit 
error (see Supplemental Material \cite{SupplMaterial}). 
If we assume that the self-broadening 
and the line shift, according to the aforementioned steady-state equations, linearly 
depend on the density, we can calculate a peak density on the order of 
\SI[parse-numbers = false]{10^{16}}{\per\centi\meter\cubed} using 
Eqs.~(\ref{eq:self-broadening}) and~ (\ref{eq:dipole-dipole_shift}).
\\
There is an apparent difference of the self-broadening and line shift between 
the two transitions of rubidium, which can be attributed to different transition dipole matrix 
elements $d_J$. While it is not possible 
to conclusively deduce any precise value for $d_J$ from our data, we calculate
ratios between the \Dline{1} and \Dline{2} broadening and shift for otherwise identical measurements, 
which are shown in Figs.~\ref{fig:broadening_shift}(a) and (b) on the right vertical axis. 
These values approach the ratios ${\sqrt{1/2}}\, d_{J,\text{\Dline{2}}}^2/d_{J,\text{\Dline{1}}}^2$ 
and $d_{J,\text{\Dline{2}}}^2/d_{J,\text{\Dline{1}}}^2$ emerging from Eqs.~(\ref{eq:self-broadening}) 
and~(\ref{eq:dipole-dipole_shift}), respectively, for large $t$ as indicated by the black triangles.
Deviations during the first $\approx\SI{10}{\nano\second}$ in the case of the self-broadening likely originate
from limited accuracy of the fits with signal wings not captured with
the scanned detuning range, asymmetries in the spectral profiles
similar to what was reported in Ref. \cite{Peyrot2018}, or asymmetries from both hyperfine splitting and velocity distribution
(see Fig.~S4, Supplemental Material \cite{SupplMaterial}).
In contrast, the measured line shift is always much smaller than
the scan range while almost vanishing for $t>\SI{8}{\nano\second}$
such that the error bars are larger than the
values themselves. Such systematic uncertainties are not properly
captured by the standard errors as derived from the employed fitting algorithms.
\par
%
%
In conclusion, we implemented a pulsed LIAD method to switch atom densities 
from $\SI[parse-numbers = false]{10^{14}}{\per\centi\metre\cubed}$ to more than
$\SI[parse-numbers = false]{10^{16}}{\per\centi\metre\cubed}$ 
on a nanosecond timescale in a micrometer-sized cell. At high densities with $nk^{-3}\approx100$
we are able to study the dipole-dipole induced self-broadening and 
Lorentz-Lorenz shift. Our measurements show 
that the interaction builds up faster than $\SI{2}{\nano\second}$, a timescale 
much shorter than the natural lifetime. The scaling between 
the \Dline{1} and \Dline{2} transition in the measurement 
supports the assumption that we observed dipolar effects in 
good agreement with the established theory.
Overall, we do not see significant transient internal dynamics
other than the one induced by the density change itself,
since the motional dephasing is the fastest timescale equilibrating
the shift and broadening of the many-body dynamics with dipolar interactions within $\approx \SI{1}{\nano\second}$. 
With a better temporal resolution (e.g., with 
superconducting single-photon detectors) and shorter desorption pulses, it will be possible 
to study the behavior of the transient dipole-dipole interaction in the first $\SI{2}{\nano\second}$, a regime 
which was not accessible in this work. The switching of the atomic medium by 
LIAD can be used with integrated photonic structures \cite{Ritter2018,Alaeian2020}, 
e.g., to realize large optical nonlinearities at a GHz bandwidth for switchable beam splitters, 
routers, and nonlinear quantum optics based on the excitation blockade.
\par
%
%
\let\oldaddcontentsline\addcontentsline
\renewcommand{\addcontentsline}[3]{}
%
\begin{acknowledgements}
The supporting data for this article are openly available from \cite{Christaller2021}. 
Additional data (e.g., raw data of the time tagger) are available on reasonable request.
\par
The authors thank Artur Skljarow for extensive support during the preparation of the final 
version of this work.
This work is supported by the Deutsche Forschungsgemeinschaft (DFG) 
via Grant No. LO 1657/7-1 under DFG SPP 1929 GiRyd.
We also gratefully acknowledge financial support by the Baden-W\"urttemberg Stiftung 
via Grant No. BWST\_ISF2019-017 under the program Internationale Spitzenforschung.
H.A. acknowledges the financial support from Eliteprogramm of Baden-W\"urttemberg Stiftung, 
Graduiertenkolleg ``Promovierte Experten für Photonische Quantentechnologien'' via 
Grant No. GRK 2642/1, and Purdue University startup grant.
C.S.A. acknowledges support from EPSRC Grant No. EP/R002061/1.
\par
F.C., M.M., and F.M. contributed equally to this work.
\end{acknowledgements}
\bibliography{2021_LIAD}
\clearpage
%
%
\onecolumngrid
	\section{Supplemental Material for \\``Transient Density-Induced Dipolar Interactions in a Thin Vapor Cell''}
\twocolumngrid

\setcounter{figure}{0}
\renewcommand{\thetable}{S\arabic{table}}
\renewcommand{\thefigure}{S\arabic{figure}}
\renewcommand{\theequation}{S\arabic{equation}}

\let\addcontentsline\oldaddcontentsline
\setcounter{secnumdepth}{3}
\tableofcontents
\section{\label{sec:background}Overview and definitions}
This Supplemental Material roughly follows the structure of the manuscript, providing additional information alongside.
In section~\ref{sec:exp_details} the experimental prerequisites to perform our LIAD measurements
are detailed and the principle data analysis procedure is introduced. Section~\ref{sec:OD_intensity}
elaborates our findings about the behavior of the desorption process
when the LIAD pulse intensity is increased. This forms a necessary foundation for the development of
our kinematic model discussed in section~\ref{sec:simulation}, as the latter makes certain assumptions
about the number of atoms which get desorbed during each pulse. Section~\ref{sec:ElecSus_fits}
frames the discussion of Fig.~\figDoneDtwo{} in the manuscript
by showing how the individual data points were processed and compared.
Section~\ref{sec:420_LIAD} concludes this work by highlighting another approach
pursued to deepen the understanding of the dipolar origin of the observed effects.\\
This Supplemental Material and the manuscript share consistent definitions of all introduced symbols.
All frequencies are given as angular frequencies, such that 
detunings $\Delta$ and decay-induced linewidths $\Gamma$ all appear on the same
scale without additional conversion factors. We define the laser detuning $\delta$ as the actively adjusted laser 
frequency $\omega_{l}$ compared to a resonant atomic transition frequency 
$\omega_{a}$ like $\delta = \omega_{l} - \omega_{a}$. 
An additional detuning occurs due to the velocity $\boldsymbol{v}$ of the atoms as we deal with a thermal gas.
Moreover, there are additional sources of line shifts such as dipole-dipole effects $\Delta_\text{dd}$ 
which are included in the perceived detuning $\Delta_\text{atom}$ of an atom given by
\begin{align}
	\Delta_\text{atom} &= \omega_{a} - \omega_{l} + \boldsymbol{k}\cdot \boldsymbol{v} + 
	\Delta_\text{dd} \\
	&= -\delta + \boldsymbol{k}\cdot \boldsymbol{v} + \Delta_\text{dd}~. \label{eq:perceived_detuning}
\end{align}
The time-of-flight $\text{TOF}(\delta)$ curves in Fig.~\figlongCell{} in the manuscript and Fig.~\ref{fig:sim_density}, 
which are given in the maps as an guide to the eye, can be understood as the time-of-flight solution
of Eq.~(\ref{eq:perceived_detuning}) for an atom with zero perceived detuning $\Delta_\text{atom}=0$ and no additional
shifts $\Delta_\text{dd}=0$ (i.e. resonant interaction under Doppler effect with otherwise
negligible linewidth) when traveling across the cell.
%
\section{\label{sec:exp_details}Experimental setup}
In order to observe a time- and detuning-resolved LIAD process and the consequent velocity and density map, 
each detected photon must be assigned 
to a time $t$ after the LIAD pulse and a detuning $\delta$. 
We use a single-photon counting module (SPCM-AQRH-10-FC from Excelitas), which
has a specified single photon timing resolution of \SI{350}{\pico\second}.
Its electrical triggers are temporally digitized
by a time tagger module (Time Tagger 20 from Swabian Instruments) with a
specified RMS (root-mean-square) jitter of $\SI{34}{\pico\second}$. This produces an absolute time\-stamp $t_{\text{SPCM},k}$
for each observed photon ($k\in\mathbb{N}$).
\begin{figure}[t]
	\includegraphics[width=\linewidth]{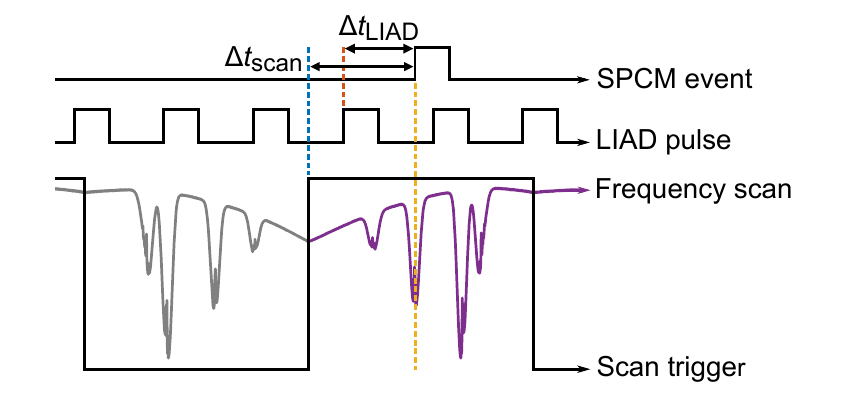}
	\caption{\label{fig:trigger_sequence} 
		Trigger sequence during the detection of a photon. 
		The time tagger module receives electrical triggers from three different sources: The SPCM, the
		LIAD pulse generator and the scan ramp generator of the probe laser. Each event
		in the SPCM channel (each photon) is referenced to the most recent LIAD pulse trigger and scan trigger.
		The resulting time differences $\Delta t_\text{LIAD}$ directly indicate the temporal position
		of the observed photon with respect to the LIAD pulse. The time differences $\Delta t_\text{scan}$
		can be converted to a temporary laser frequency by comparing the scan trigger sequence 
		to an absolute frequency reference indicated by an atomic saturation spectroscopy signal (purple curve).
	}
\end{figure}
\\
The LIAD laser pulses are electrically triggered from a pulse generator (Bergmann BME\_SG08p)
at \SI{50}{\kilo\hertz} repetition rate
with a specified delay resolution of \SI{25}{\pico\second} and an output to output RMS jitter of
less than \SI{50}{\pico\second}. These electrical pulses are fed into a Q-switched
\SI{532}{\nano\meter} laser (BrightSolutions WedgeHF 532) with an experimentally determined
pulse-to-pulse jitter of less than \SI{500}{\pico\second}. We use an additional synchronized
output from the pulse generator
as temporal reference for the laser pulses generating additional timestamps $t_{\text{LIAD},\ell}, (\ell\in\mathbb{N})$. \\
At the same time, a Littrow-configuration, external-cavity diode laser (Toptica DL Pro) is
continuously frequency-scanned over the ranges shown in the figures of the manuscript at a rate of \SI{11}{\hertz}. This ensures
that the effective detuning $\delta$ is quasi constant during each individual
LIAD pulse and the transient regime, which is on the order of less than a microsecond. It additionally provides a large number
of LIAD pulses for each detuning in every dataset spanning over an integration time
in the order of one day. The scan ramp triggers are fed into the time tagger module and generate
a series of timestamps $t_{\text{scan},m}, (m\in\mathbb{N})$.
This overall trigger sequence for a single photon detection 
is schematically depicted in Fig.~\ref{fig:trigger_sequence}.\\
During analysis, time differences 
\begin{align}
	\Delta t_{\text{LIAD},k} &=t_{\text{SPCM},k} - t_{\text{LIAD},\ell} \quad\text{and}\\
	\Delta t_{\text{scan},k} &=t_{\text{SPCM},k} - t_{\text{scan},m}
\end{align}
are calculated, such that each SPCM time\-stamp gets assigned to the most recent LIAD and scan trigger.
The value $\Delta t_{\text{LIAD},k}$ reflects the timing with respect to the desorption pulse.
The different electrical and optical propagation times are corrected by observing LIAD pulses
directly on the same SPCM and subtracting their average peak pulse position to obtain time differences on
a time axis $t$, which is plotted in all figures in this work.
\\
Similarly, $\Delta t_{\text{scan},k}$ relates to the temporary probe laser frequency $\omega_{\text{l},k}$ by a certain
function $\omega_{l}(\Delta t_{\text{scan}}, t_\text{SPCM})$. The direct dependency on the absolute time
$t_\text{SPCM}$ indicates that laser frequency drifts must be corrected.
This is achieved by calibrating to an ultra low expansion (ULE) cavity
captured continuously on a digital oscilloscope. The ULE cavity 
has a free spectral range of $2\pi\times\SI{1.5}{\giga\hertz}$ and an overall negligible drift. 
The absolute frequency with respect to the atomic transitions can be determined by additionally
observing a saturated rubidium spectrum on the same oscilloscope. All values $\delta$
are given with respect to the center-of-mass of the shown transition(s). Note that 
the function $\omega_{l}(\Delta t_{\text{scan}}, t_\text{SPCM})$ is calculated stepwise to
reflect the different behavior during the rising and falling scan ramp.
We estimate an overall statistical uncertainty for $\delta$ 
of less than $2\pi\times\SI{10}{\mega\hertz}$, mainly limited by fluctuations of the scan rate 
during each scan ramp not captured by the $2\pi\times\SI{1.5}{\giga\hertz}$ ULE cavity.
\\
The plotted maps with respect to $t$ and $\delta$ are calculated by binning relevant
pairs $(t, \delta)_k$ into a 2D histogram.
A bin width of $2\pi\times\SI{10}{\mega\hertz}$ to $2\pi\times\SI{50}{\mega\hertz}$
for $\delta$ and $\SI{500}{\pico\second}$ 
for $t$ is used throughout this work. The bin width of $t$ is thus chosen on the same order of magnitude as 
the time jitter limiting the overall measurement. The binning of the probe laser detuning 
can be chosen more arbitrarily to suppress the noise. 
For the displayed data we use a bin-width larger than the free-running linewidth 
of the probe laser ($\nu \approx 2\pi\times\SI{800}{\kilo\hertz}$).
\\
From the various measurements shown in the manuscript and the experience with our setup, we can state, 
that the observations due to the LIAD effect are repeatable after months and years. Yet we 
have to mention, that during measurements with high LIAD pulse intensity, the 
cell is locally modified. This modification manifests itself as a decreased transmission
or visibly brown spot, which, depending on the total exposure,
can be either temporary and healed by uniform heating or stay permanently.
We attribute this behavior to an alteration of the sapphire coating
after bombardment with rubidium atoms. We could not observe
such a change in transmission in a similar cell filled with air under otherwise
identical conditions. We also note that a similar discoloration and
loss of transmission is a well-known effect for cells filled
with alkali vapor at elevated temperatures \cite{Sekiguchi2017}.
It could be possible that the LIAD effect and velocity distribution
locally produce a similar phenomenon even at lower temperatures.
To minimize the impact of such effects on our measurements,
we monitor the probe transmission and regularly move the cell to a spot with the 
same cell thickness to continue the measurement once
a noticeable transmission loss is observed. Thereby, we gain reproducible results
over multiple measurement runs.
%
\section{\label{sec:OD_intensity}Atomic density for increasing LIAD intensity}
\begin{figure}[t]
	\includegraphics[width=\linewidth]{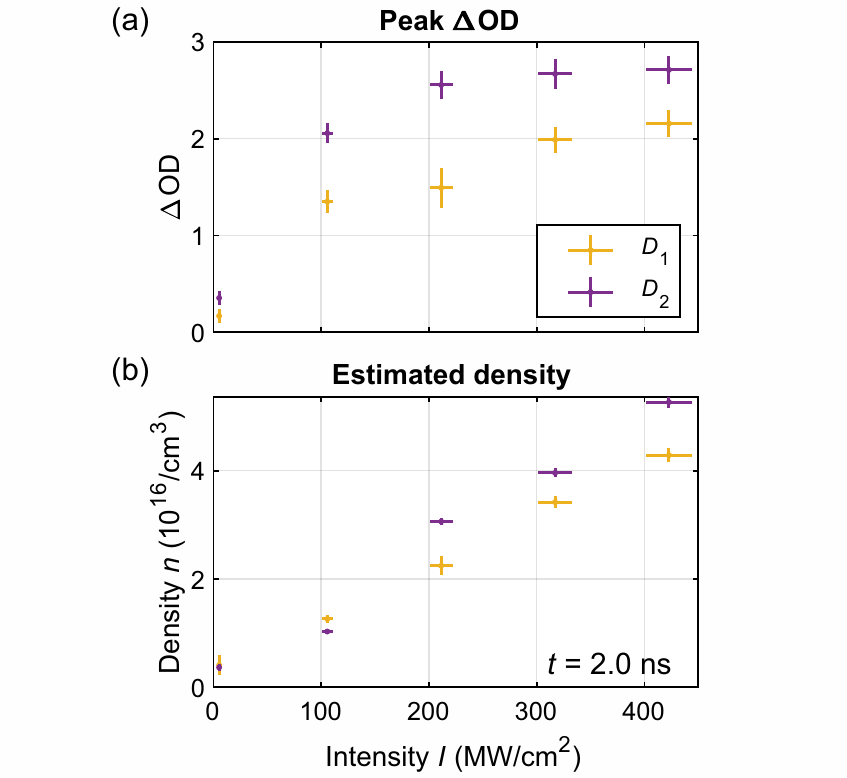}
	\caption{\label{fig:OD_density} 
		(a) Measured values of $\Delta \text{OD}_\text{peak}$ for various
		LIAD pulse peak intensities. All other parameters of the system 
		are kept identical and comparable to the settings discussed in Fig.~\figshortCell{} in the manuscript.
		(b) Density $n(t=\SI{2}{\nano\second})$ calculated from the fitted broadening and shift values using 
		Eqs.~(\eqselfbroadening{}) and~(\eqdipoledipoleshift{}) from the manuscript.
	}
\end{figure}

In this work we present the change of the optical depth $\Delta \text{OD}$ as an indicator for the
temporally-controlled, additional atomic density (or equivalently the number of desorbed atoms) 
compared to a vapor cell homogeneously heated to a certain background temperature. 
It is plausible to assume that for LIAD pulses of fixed duration, the number of desorbed atoms increases 
with increasing LIAD pulse peak intensity $I$, since the total energy introduced into the system increases. 
While we could not deduce a simple relationship between the parameters for 
a velocity distribution model (see section~\ref{sec:simulation}) and the LIAD pulse parameters, 
we still observe well-defined relationships between $I$ and 
key features of the corresponding $\Delta \text{OD}$ maps.
\\
One such feature is the peak optical depth value reached on resonance, $\Delta \text{OD}_\text{peak}$, as an indicator 
for the total number of atoms desorbed during the process. Fig.~\ref{fig:OD_density}(a) shows that these values 
monotonically increase for both the \Dline{1} and \Dline{2} transition with increasing $I$. As the 
attainable atomic density is independent of the probe wavelength and therefore roughly identical 
in both cases, the ratio between the individual $\Delta \text{OD}_\text{peak}$ values for 
both transitions should be proportional to their resonant scattering cross section $\sigma_0$, 
which in turn can be related to the transition dipole moment $d$ as 
$\sigma_0\propto|d|^2$ \cite{Loudon2000}. This assumes steady-state behavior 
and is therefore not perfectly reproduced in our case. A much stronger absorption of light at 
the \Dline{2} transition however is visible, which is expected given the transition dipole moment squared 
is larger by factor of roughly 2 compared to the \Dline{1} transition \cite{SteckRb}.  
The displayed horizontal error bars in Fig.~\ref{fig:OD_density} 
show the $\SI{10}{\percent}$ standard deviation 
of the LIAD peak intensity due to power fluctuations. The vertical error bar
of the $\Delta \text{OD}$ is determined via error propagation from the statistical uncertainty of 
the photon counts $N_\text{photon}$, which is given by 
$\Delta N_\text{photon} = \sqrt{N_\text{photon}}$. 
\\
The highest $\Delta \text{OD}_\text{peak}$ shown in Fig.~\ref{fig:OD_density}(a) is at 
the limit of what can be detected with our setup due to the low number of events detected by the SPCM modules compared 
to the background noise of the system (dark count rate $R_\text{dark}$). From the datasheet provided by 
the manufacturer and reference measurements, we estimate at $R_\text{dark}\approx\SI{1.5e3}{\per\second}$ 
with an average count rate $R_\text{avg}\approx\SI{1e6}{\per\second}$, an overall 
$\text{OD}_\text{peak}=\ln(R_\text{avg}/R_\text{dark})\approx 6.5$ might just be accessible.
This also includes the optical depth contribution of the thermal background vapor, which
was less than $0.5$ for the largest cell thicknesses presented in this work ($L=\SI{6.24}{\micro\meter}$).
The observed behavior $\Delta \text{OD}_\text{peak}$ with $I$ could therefore be attributed 
to statistical limits of the SPCM in addition to the density broadening and the unknown microscopic behavior 
of the cell wall's material. The latter include temporary (i.e. debris, accumulation of rubidium atoms) and 
permanent (i.e. cell damage) modifications for prolonged measurement cycles at one spot of the wedge-shaped cell.
\\
The relation between a growing $\Delta \text{OD}_\text{peak}$ and the underlying atomic density $n_\text{peak}$ 
becomes nonlinear at high densities due to the atomic interactions and cannot be
calculated directly without precise knowledge about the
density broadening effect and the actual density- and velocity distribution. 
The actual peak density $n_\text{peak}$ might therefore show 
e.g. a linear trend with $I$, while the growth of $\Delta \text{OD}_\text{peak}$ 
appears to saturate [Fig.~\ref{fig:OD_density}(a) data points at larger $I$]. 
The manuscript mentions a possible approach using steady-state derivations starting 
from the fitted broadening and shift. There, the presented order of magnitude of the density is calculated using 
Eqs.~(\eqselfbroadening{}) and~(\eqdipoledipoleshift{}) from the manuscript. As the broadening and the shift lead
to almost the same density values, we can calculate the average density from these two values and plot 
the estimated density at $t=\SI{2}{\nano\second}$ for different LIAD peak intensities, 
as shown in Fig.~\ref{fig:OD_density}(b). There, we observe an almost linear behavior between 
the LIAD peak intensity and the estimated density from both the \Dline{1} and \Dline{2} transition. 
The uncertainties of the estimated density result from the susceptibility fits done with \textsc{ElecSus}. 
Note, that we apply a steady-state model to determine these density values and we have no other independent 
way to measure the transient density (see section~\ref{sec:ElecSus_fits}).
%
\section{\label{sec:simulation}Kinematic model and simulation}
\begin{figure}[t]
	\includegraphics[width=\linewidth]{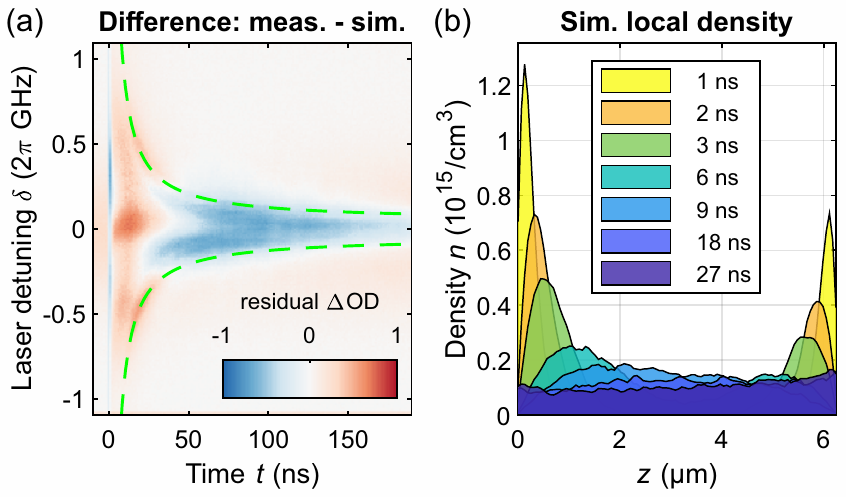}
	\caption{\label{fig:sim_density}
		(a) Difference between measured and simulated $\Delta \text{OD}$ map, which are shown 
		in Fig.~\figlongCell{} in the manuscript. The residual $\Delta \text{OD}$ value indicates 
		where the simulation under- or overestimates the measurement. The $\text{TOF}(\delta)$ 
		is shown with two dashed green lines.
		(b) Simulated local density in the $\SI{6.24}{\micro\metre}$ thick cell. The high density 
		at both cell walls ($z=\SI{0}{\micro\metre}$ and $z=\SI{6.24}{\micro\metre}$) is decreasing 
		with increasing time after the LIAD pulse. After $t \approx \SI{18}{\nano\second}$ the atoms 
		are uniformly distributed along the $z$ axis of the cell.
	}
\end{figure}

This section describes how the kinematic model and the Monte Carlo simulation are set up to generate the 
optical depth map numerically [Fig.~\figlongCell{}(b) in the manuscript].
\\
As a first step, we pick atoms on one of the two inner cell walls ($z=0$ or $z=L$, with the cell thickness $L$). 
The $x$ and $y$ position of the atoms are distributed normally, where the width of the distribution is given 
by the waist radius of the LIAD beam $w_\text{LIAD} = \SI{13.7}{\micro\meter}$ 
(atoms are only desorbed where the LIAD beam hits the cell). 
We choose a velocity distribution of the form $f(v,\varphi,\theta) = \left[4/(\sqrt{\pi}a^3)v^2\exp{(-v^2/a^2)}\right] \cos(\theta)$ 
with the parameter $a$, adjusted to agree with the measurement. As for the angular part of this distribution, 
the azimuthal angle $\varphi$ is uniformly distributed and the polar angle $\theta$ is distributed 
according to the $\cos(\theta)$-Knudsen law \cite{Knudsen1934,Comsa1985}.
The desorption time of the atoms is defined by the temporal shape of the LIAD pulse. This pulse has a shape 
similar to the Blackman window with a length of $\SI{1.1}{\nano\second}$ (FWHM). 
The asymmetry of the measurement between the two atom clouds moving in or against the laser propagation direction 
(atoms with positive and negative detuning, respectively) is captured via a ratio variable $r$. 
This ratio $r$ is defined as the total number of atoms with negative detuning over the total number of atoms 
with positive detuning. After each atom traveled through the cell and hit the opposite wall, it can lead to 
a re-emission of another atom from the cell wall with a certain probability $p_{r}$ and with a new velocity 
($v$: Maxwell-Boltzmann; $\varphi, \theta$: uniform). Only atoms which enter the cylindrical probe beam 
with a radius of $w_\text{probe} = \SI{2.0}{\micro\meter}$ are considered. 
The actual transversal Gaussian profile of the probe beam is therefore approximated by
a tophat profile to simplify the decision whether a particle is currently
inside the probe beam. This is valid due to its small diameter compared to the LIAD
beam and allows for an ad-hoc implementation of transit-time broadening effects
without the need to solve the time-dependent optical Bloch equations for each atom.
\begin{table}[b]
	\caption{\label{tab:model_parameter}%
		Parameters used in the simulation to reproduce the measurement. The values are 
		adjusted to minimize the RMS of the residual $\Delta\text{OD}$ map, shown in 
		Fig.~\ref{fig:sim_density}(a).
	}
	\begin{ruledtabular}
		\begin{tabular}{llll}
			variable & $a$ & $r$ & $p_{r}$ \\
			\colrule\vspace*{-0.95em}\\
			value & \SI{271}{\metre\per\second} & \num{0.6} & \num{0.84} \\
		\end{tabular}
	\end{ruledtabular}
\end{table}
The simulation parameters for Fig.~\figlongCell{}(b) in the manuscript  
are displayed in Table~\ref{tab:model_parameter}. 
These values were simultaneously optimized using a supervised pattern search algorithm by 
minimizing the root-mean-square (RMS) value of the residual $\Delta \text{OD}$ map. This map 
is the difference between the measurement and the simulation, shown in Fig.~\ref{fig:sim_density}(a).
From the parameter $a$ one can estimate a temperature corresponding to the desorbed atoms, which is 
$T_\text{sim} \approx \SI{100}{\degreeCelsius}$.
\\
The second step is the calculation of the time- and $z$-dependent local density $n(t,z)$ of the atoms. 
The $z$ axis with the cell thickness $L$ is sliced into $N_z$ slices. For each simulation time step 
and each slice, the number of atoms in a slice is counted and divided by the slice volume.
For large numbers of atoms it is possible to only simulate a fraction of all atoms and rescale this value
accordingly. This gives a time- and $z$-dependent estimation of the density in the cell, 
shown in Fig.~\ref{fig:sim_density}(b).
\\
As a third step, the scattering cross section for the simulated time steps and laser detunings 
are calculated using a steady-state model as an approximation. For each 
time step $t$, laser detuning $\delta$ and atom $i$ 
the resulting detuning is determined as $\Delta_\text{atom} = -\delta + k v_{z,i} + \Delta_{\text{dd},i}$, 
where $k = 2\pi/\lambda$ is the laser wave number and $\Delta_\text{dd}$ is the additional dipole-dipole line shift, 
which depends on the local density.  
In addition to the natural decay rate $\Gamma_0$, we also include the transit broadening $1/\tau_\text{tt}$ 
for each individual atom to model the finite probe size effect. 
After including the density-dependent self-broadening $\Gamma_\text{self}$, the total broadening reads 
as $\Gamma = \Gamma_0 + 1/\tau_{\text{tt},i} + \Gamma_{\text{self},i}$. 
\\
\begin{figure*}[t]
	\includegraphics[width=\linewidth]{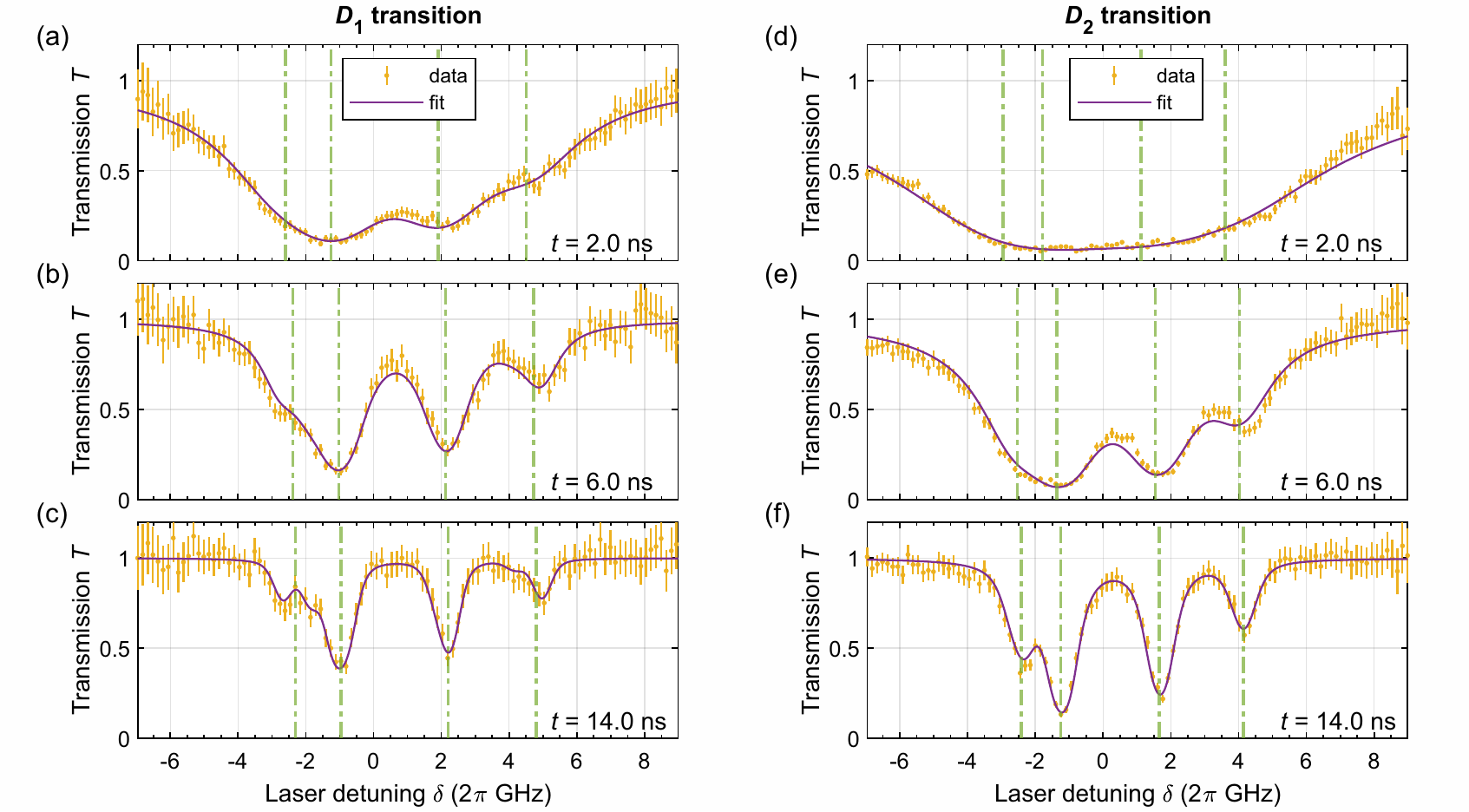}
	\caption{\label{fig:ElecSus_fit} 
		Fits of an electric susceptibility model to the measured data 
		of the \Dline{1} transition in (a,b,c) and \Dline{2} transition in (d,e,f).
		The data points (yellow, only every third data point shown) 
		and the fit (purple) are in good agreement. The fits are performed for every time $t$ in steps 
		of the time binning. Exemplary data are the times $t=\SI{2}{\nano\second}$ in (a,d), 
		$t=\SI{8}{\nano\second}$ in (b,e), and $t=\SI{14}{\nano\second}$ in (c,f). 
		For the traceability of the changing line shift over time, we mark the shifted frequencies 
		of the two ground state hyperfine splittings of the two rubidium isotopes 
		with four green vertical lines.
		The cell thickness is 
		$L = \SI{0.78 \pm 0.02}{\micro\metre}$, the peak intensity of the LIAD pulse is 
		$I = \SI{317\pm32}{\mega\watt\per\square\centi\metre}$, and the reservoir temperature is 
		$T_\text{res} \approx \SI{180}{\degreeCelsius}$.
	}
\end{figure*}
The steady-state scattering cross section 
is defined as~\cite{SteckRb}
\begin{equation}
	\sigma(\Delta_\text{atom},I) = \frac{\sigma_0}{1+4(\Delta_\text{atom}/\Gamma_0)^2+(I/I_\text{sat})}~, 
\end{equation}
where $\sigma_0$ is the resonant scattering cross section, defined as~\cite{SteckRb}
\begin{equation}
	\sigma_0 = \frac{\hbar \omega_{a} \Gamma_0}{2 I_\text{sat}}~.
\end{equation}
With the additional broadening effects the scattering cross section has to be normalized with $\Gamma_0/\Gamma$. 
As we are in the weak probe regime with $I_\text{probe}<I_\text{sat}$, 
the scattering cross section can be written as 
\begin{equation}
	\sigma(\Delta_\text{atom},\Gamma) = \frac{\Gamma_0}{\Gamma}\frac{\sigma_0}{1+4(\Delta_\text{atom}/\Gamma)^2}~.
\end{equation}
To capture the dipolar dynamics 
of the atom-probe interaction, we use time-dependent $\Delta_\text{dd}$ and $\Gamma_\text{self}$ values 
due to the time- and $z$-dependent atomic density [see Eqs.~(\eqselfbroadening{}) and~(\eqdipoledipoleshift{}) in the manuscript]. 
The scattering cross sections are accumulated and normalized for all simulated atoms within the probe region 
for each time step, laser detuning and $z$ slice as $\sigma(t,\delta,z)$. This can be understood as the
average cross section $\sigma(\delta)$ contributed by an atom found at time $t$ at location $z$.
\\
The last step is the conversion of the calculated density $n(t,z)$ and scattering 
cross section $\sigma(t,\delta,z)$ to an optical depth according to the Beer-Lambert law.
This correctly captures shadowing effects among atoms if the density in each $z$ slice is low
compared to its thickness (s.t. $n\sigma L/N_z\ll1$)
and can be formulated for $z$-dependent $n$ and $\sigma$.
Since we only simulate the desorbed atoms without any background gas, the 
calculated change of the optical depth is
\begin{equation}
	\Delta \text{OD}(t,\delta) = \int_0^L \sigma(t,\delta,z) n(t,z) \mathrm{d}z~.
	\label{eq:DeltaOD_sim}
\end{equation}
%
\section{\label{sec:ElecSus_fits}\textsc{ElecSus} fitting procedure}
Here we show some exemplary electric susceptibility fits done with \textsc{ElecSus} \cite{Zentile2015}. The data points 
were weighted according to their inverse uncertainties shown in Fig.~\ref{fig:ElecSus_fit}. The error bars on 
the wings of the spectrum are larger due to a filter etalon in the setup, reducing the number of 
photon counts $N_\text{photon}$ and thereby increasing the statistical uncertainty 
$\Delta N_\text{photon} = \sqrt{N_\text{photon}}$. We start fitting from larger times $t$, with negligible interaction, 
to shorter times, showing strong interaction effects, 
using the results from the previous fit as initial parameters for the next fit. 
Note that the \textsc{ElecSus} internal treatment of the self-broadening 
is manually switched off to get the full information of the broadening from the fits. 
We also set the Doppler temperature to $T_\text{Doppler} = \SI{0}{\degreeCelsius}$, as we mainly want 
a Lorentzian profile, which is in first approximation justified to capture the self-broadening 
($\Gamma_\text{self} > \Gamma_\text{Doppler}$). The free fit parameters for our \textsc{ElecSus} fits 
are the width of the Lorentzian profile ($\Gamma_\text{buf}$, normally used for a buffer gas broadening), 
the line shift and the temperature of the cell $T_\text{cell}$, which is used to calculate a temperature-dependent 
atomic density. Note, that this intrinsic temperature-dependent density in \textsc{ElecSus} is not a useful quantity
in our anisotropic system and therefore not considered in this work. In our system the velocity of the atoms
has a certain direction, where a possible velocity distribution is discussed in our kinematic model 
(see also Section~\ref{sec:simulation}).
\begin{figure}[t]
	\includegraphics[width=\linewidth]{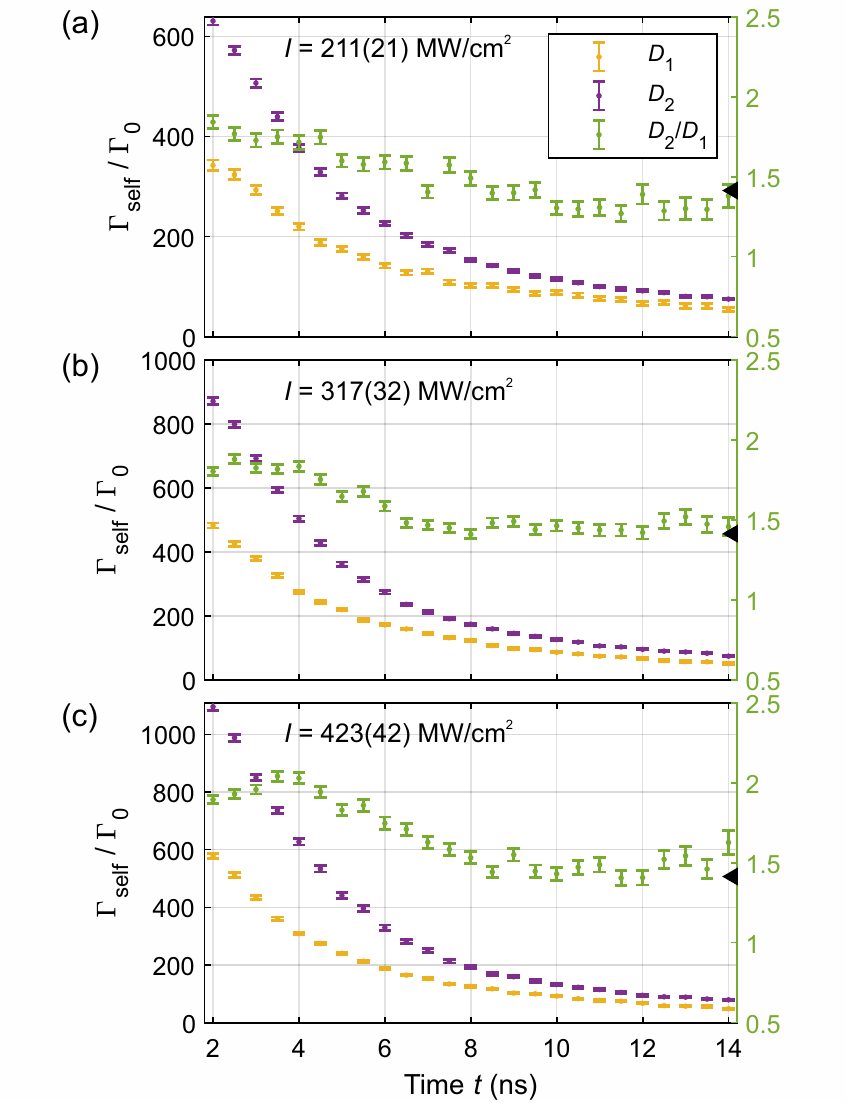}
	\caption{\label{fig:broadening_with_ratio} 
		Self-broadening of the \Dline{1} (yellow) and \Dline{2} (purple) transition over time 
		for three different LIAD intensities $I$. Additionally, the ratio \Dline{2}/\Dline{1} (green) 
		is plotted, which is approaching the theoretical steady-state ratio (black triangle). 
		For high intensities, e.g. in (c), the broadening of the \Dline{2} transition is overestimated 
		during the fit (see also Fig.~\ref{fig:ElecSus_fit}), which is one reason for the deviation of the ratio from the theoretical value in the first $\approx\SI{10}{\nano\second}$.
	}
\end{figure}
\\
The \textsc{ElecSus} software calculates the $1\sigma$ standard error of the fitted variables.
These standard errors 
are plotted as vertical error bars in Fig.~\figDoneDtwo{} in the manuscript 
and in Fig.~\ref{fig:broadening_with_ratio}.
There are no horizontal error bars shown in both these figures, as 
they originate from histogram binning. In this sense, the horizontal precision
is limited by the bin width which is motivated by the
the total temporal precision $\Delta t_\text{total} = \SI{850}{\pico\second}$
of each captured event.
This $\Delta t_\text{total}$ is the
sum of the LIAD pulse-to-pulse jitter and the time resolution of the SPCM,
which limit the bandwidth of the measurement.
The goodness of each fit is evaluated by calculating the normalized root-mean-square deviation
$\mathrm{NRMSD}$ as given in the manuscript,
which is calculated from the root-mean-square deviation $\mathrm{RMSD} = N^{-1}\sqrt{\sum_{i=1}^N(T_i-\hat{T}_i)^2}$
between the measured transmission points $T_i$ and the fitted transmission value $\hat{T}_i$.
These are then normalized using the minimum and maximum transmission values $T_\text{min}=0,T_\text{max}=1$
as $\mathrm{NRMSD}=\mathrm{RMSD}/(T_\text{max}-T_\text{min})$. The obtained values are similar for both probe
transitions and typically at $\mathrm{NRMSD}\approx\SI{5}{\percent}$.
\\
The electric susceptibility model is a steady-state solution of the atom-light interaction. 
As mentioned in 
the manuscript, we cannot properly fit the data points in the first $\SI{2}{\nano\second}$. 
Initially the broadening is large and the signals' wings are not properly captured
with $\approx2\pi\times\SI{16}{\giga\hertz}$ scan range in our experiment. Also 
the LIAD pulse, which is present until $t\approx \SI{1.5}{\nano\second}$, distorts the measured data.
There are several effects which could systematically distort the
quality of the fit results or render the model unsuitable for
later times $t$
(ordered by the hypothesized contribution from strongest to lowest according to the authors):
A non-isotropic velocity distribution as produced by the LIAD pulse,
geometry-dependent effects like the collective Lamb shift,
mathematical artifacts due to a clipped detuning range,
asymmetries from different relative hyperfine transition strengths and dipole moments,
asymmetric line shapes caused by any other effect
(e.g. surface potentials),
multi-particle interactions,
differences in the experimental configuration between both transitions,
and temporally transient dipolar effects.
\\
The ratio between the line shifts can be properly recovered from the fits for $t<\SI{7}{\nano\second}$
even if the wings are not fully captured, as the the dipole-dipole shift is smaller than the
self-broadening and can be measured as an absolute offset instead of a line width.
For very large times, corresponding
to low densities, the ratio between
the vanishing shifts on the \Dline{1} and \Dline{2} transition will 
be very susceptible to any residual offset. In
this region, which occurs at different times for self-broadening and line shift
due to their different absolute values,
we observe varying and unstable behavior. We suspect that a combination
of all these effects causes deviations from the theoretical
ratios between the \Dline{1} and \Dline{2} transition as reported in Fig.~\figDoneDtwo{} in the manuscript.
\begin{figure}[b]
	\includegraphics[width=\linewidth]{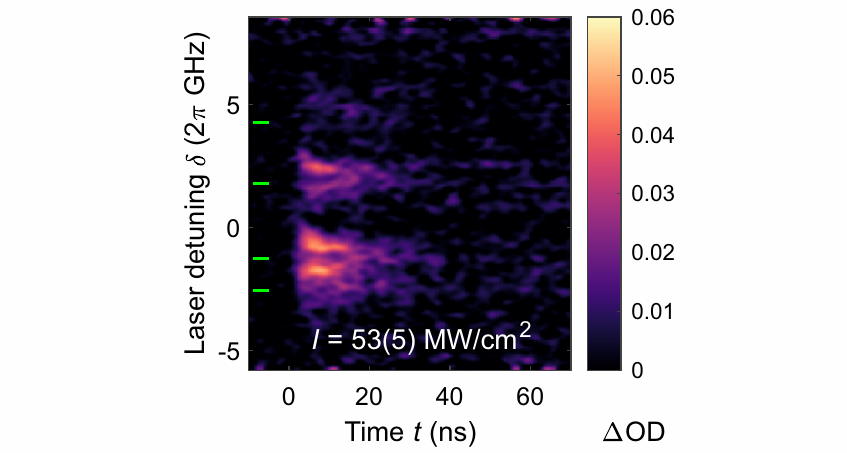}
	\caption{\label{fig:420_LIAD} Measured $\Delta \text{OD}$ map on the $5S_{1/2} \to 6P_{1/2}$ 
		transition at a cell thickness of \SI{2.34\pm0.04}{\micro\meter} and a reservoir temperature of 
		$T_\text{res} \approx \SI{230}{\degreeCelsius}$. The four green markers indicate 
		the ground state hyperfine splitting of the two isotopes of rubidium, respectively. 
		Note, that the image has been
		filtered using a Gaussian filter to reduce the overall noise in the data. 
		The intensity $I$ is the peak intensity of the LIAD pulse.
	}
\end{figure}
\\
While the self-broadening is directly visible in Fig.~\ref{fig:ElecSus_fit} for exemplary times, 
we also show the 
fitted line shift by drawing the frequencies of the two ground state hyperfine splittings 
of the two rubidium isotopes, to emphasize the changing line shift over time.
The figure also highlights that artifacts due to a clipped laser detuning range
would impact the \Dline{1} transition less (or at different times) than the \Dline{2} transition.
This idea is supported by the reduced variation in the \Dline{2}/\Dline{1} ratio in
Fig.~\ref{fig:broadening_with_ratio}(a), which indeed has the lowest overall density and broadening.
\\
To underline our observation of the transient density-dependent dipolar interactions, 
we show the time evolution of the self-broadening for two additional LIAD intensities in 
Fig.~\ref{fig:broadening_with_ratio}. With increasing intensity, the density is also increasing,  
leading to a larger self-broadening. The ratio of \Dline{2}/\Dline{1} shows a similar behavior for 
all intensities.
%
\section{\label{sec:420_LIAD}Pulsed LIAD probed at 420\,\lowercase{nm}}
The experiments in the manuscript were performed using probe lasers at the \Dline{1} and \Dline{2} 
ground state transitions, with relatively large transition dipole moments. 
This, however, induces 
limits considering the accessible density regimes before broadening effects become larger than the
captured frequency range as discussed above. It can be overcome by probing on a transition with a lower
transition dipole moment.
\\
We therefore modified our measurement setup to include a \SI{421.6}{\nano\meter} laser probing 
the $5S_{1/2} \to 6P_{1/2}$ transition, where the squared transition dipole moments 
are reduced by roughly two orders of magnitude \cite{Sibalic2017}. The measured data for total integration times 
similar to the previously discussed cases exhibits a much noisier signal and lower $\Delta \text{OD}$, 
which is expected due to the much weaker transition dipole moment. Consequently, the resulting map shown in 
Fig.~\ref{fig:420_LIAD} could only be captured for relatively strong LIAD pulses, a thicker part of the cell, 
and a higher reservoir and cell temperature. Important features discussed in the manuscript 
are reproduced. Both signals representing atoms moving towards and away from the laser 
are visible and exhibit the same asymmetry towards atoms moving in propagation direction of the LIAD laser. 
We were not able to measure a line shift or broadening which is attributed to 
the significantly weaker transition dipole moment. 
This further supports the conclusion that the observed effects in the manuscript indeed have a dipolar origin.
%
\par
F.C., M.M., and F.M. contributed equally to this work.
\end{document}